\documentstyle[12pt,thmsa,sw20lart]{article}
\input tcilatex
\begin{document}

\author{I. Radinschi\thanks{%
iradinsc@phys.tuiasi.ro}}
\date{Department of Physics, ``Gh. Asachi'' Technical University, Iasi, 6600,
Romania}
\title{M\o ller Energy-Momentum Complex for an Axially Symmetric Scalar Field}
\maketitle

\begin{abstract}
We calculate the energy-distribution for an axially symmetric scalar field
in the M\o ller prescription. The total energy is given by the parameter $m$
of the space-time.


\end{abstract}

\section{Introduction}

The subject of energy-momentum localization in general relativity continues
to be one of the most intricate because there is no given yet a generally
accepted expression for the energy and momentum. It is well-known that
various energy-momentum complexes [1]-[17] can give the same energy
distribution for a given space-time. Recently, Virbhadra [10] studied if it
is possible to obtain the same expression of the energy in the case of a
Kerr-Schild metric by using the energy-momentum complexes of Einstein [18],
Landau and Lifshtz [19], Papapetrou [20] and Weinberg [21] ELLPW. He
concluded that these definitions lead to the same result. On the other hand,
these definitions disagree for the most general nonstatic spherically
symmetric metric [10]. Only the Einstein energy-momentum complex gives the
same expression for the energy distribution when the calculations are
performed in the Kerr-Schild Cartesian and Schwarzschild Cartesian
coordinates.

Also, many results recently obtained, [16], [17], [22], [23] demonstrated
that the M\o ller energy-momentum complex [24] is a good tool for obtaining
the energy distribution of a given space-time. M\o ller energy-momentum
complex allows to make the calculations in any coordinate system. In his
recent paper, Lessner [25] concluded that the M\o ller definition is a
powerful concept of energy and momentum in general relativity. Very
interesting is the Cooperstock [26] hypothesis. He sustain that the energy
and momentum are confined to the regions of nonvanishing energy-momentum
tensor of the matter and all non-gravitational fields.

Also, Chang, Nester and Chen [9] showed that the energy-momentum complexes
are actually quasilocal and legitimate expressions for the energy-momentum.

The purpose of this paper is to compute the energy distribution, for an
axially symmetric solution that describes the space-time, which is endowed
with a scalar field, in the M\o ller prescription. Recently gravitational
lensing due to this space-time has been studied in a great detail [27],
[28]. We use the geometrized units $(G=1$, $c=1)$ and follow the convention
that Latin indices run from $0$ to $3$.

\section{M\o ller Energy-Momentum Complex for on Axially \break
Symmetric
Scalar Field}

It is well-known that the Kaluza-Klein and the superstring theories predict
the scalar fields as a fundamental interaction in physics. Scalar fields are
fundamental components of the Brans-Dicke theory and of the inflationary
models. Also, they are a good candidate for the dark matter in spiral
galaxies. Because they interact very weakly with mater we have never seen
one but many of the theories containing scalar fields are in good
concordance with measurements in weak gravitational fields. Also, we expect
that they can play an important role in strong gravitational fields like at
the origin of the universe or in pulsars or black holes. The metric that we
consider [27], [28] is an axially symmetric solution to the field equations
derived from the action for gravity minimally coupled to a scalar field. The
solution is 
\begin{equation}
ds^{2}=\left( 1-\frac{2m}{r}\right) dt^{2}-\frac{e^{2k_{a}}dr^{2}}{1-\frac{2m%
}{r}}-r^{2}\left( e^{2k_{a}}d\theta ^{2}+\sin ^{2}\theta d\varphi
^{2}\right) ,  \tag{1}
\end{equation}
with 
\begin{equation}
e^{2k_{a}}=\left( 1+\frac{m^{2}\sin ^{2}\theta }{r^{2}\left( 1-\frac{2m}{r}%
\right) }\right) ^{-1/a^{2}},  \tag{2}
\end{equation}
\[
\phi =\frac{1}{2a}\ln \left( 1-\frac{2m}{r}\right) , 
\]
where $a$ is a constant of integration and $\phi $ is the scalar field. This
solution is one of the new classes of solutions to the Einstein-Maxwell
theory non minimally coupled to a dilatonic field [28]. The metric given by
(1) is almost spherically symmetric and represents a gravitational body
(gravitational monopole) with scalar field. The scalar field deforms the
spherically symmetry. We observe, that when $a\rightarrow \infty $, we
recover the Schwarzschild solution. This metric can be employed to model the
exterior field of a macroscopic object endowed with a minimally coupled
scalar field, and it can be matched to a regular interior solution.

The M\o ller energy-momentum complex [24] is given by 
\begin{equation}
\theta _i^{\;k}=\frac 1{8\pi }\frac{\partial \chi _i^{\;kl}}{\partial x^l} 
\tag{3}
\end{equation}
where 
\begin{equation}
\chi _i^{\;kl}=\sqrt{-g}\left( \frac{\partial g_{in}}{\partial x^m}-\frac{%
\partial g_{im}}{\partial x^n}\right) g^{km}g^{en}.  \tag{4}
\end{equation}
The energy in the M\o ller prescription is given by 
\begin{equation}
E=\iiint \theta _0^{\;0}dx^1dx^2dx^3=\frac 1{8\pi }\iiint \frac{\partial
\chi _0^{\;0}}{\partial \chi ^l}dx^1dx^2dx^3.  \tag{5}
\end{equation}

The M\o ller energy-momentum complex is not necessary to carry out the
calculations in the quasi-Cartesian coordinates, so we can calculate in the
spherical coordinates.

For the metric given by (1) the only required component of $\chi _{0}^{\;0l}$
is 
\begin{equation}
\chi _{0}^{\;01}=2m\sin \theta .  \tag{6}
\end{equation}

Plugging (6) in (5) and applying the Gauss theorem we obtain 
\begin{equation}
E=m.  \tag{7}
\end{equation}

The energy distribution is given by the mass $m$.

\section{Discussion}

We use an axially symmetric solution to the field equations for a scalar
field minimally coupled to gravity. Virbhadra et al. [23] demonstrated that
the presence of on object endowed with a scalar field acting as a
gravitational lens would manifest through new lensing configurations. For
the above situation we obtain the energy for the metric given by (1) by
using the M\o ller energy-momentum complex. The energy distribution is given
by the mass $m$. In the case of the Schwarzschild metric we obtain the same
result. The Schwarzschild solution is a particular case of the metric that
we use, when we have $a\rightarrow \infty $.

\end{document}